\documentclass{moriond}

\usepackage{amsmath}
\usepackage{amssymb}
\usepackage{enumerate}

\def\be{\begin{equation}}
\def\ee{\end{equation}}
\def\bea{\begin{eqnarray}}
\def\eea{\end{eqnarray}}

\newcommand{\mys}[1]{%
    \scriptscriptstyle{#1}
    }

\newcommand{\Photo}{\includegraphics[height=35mm]{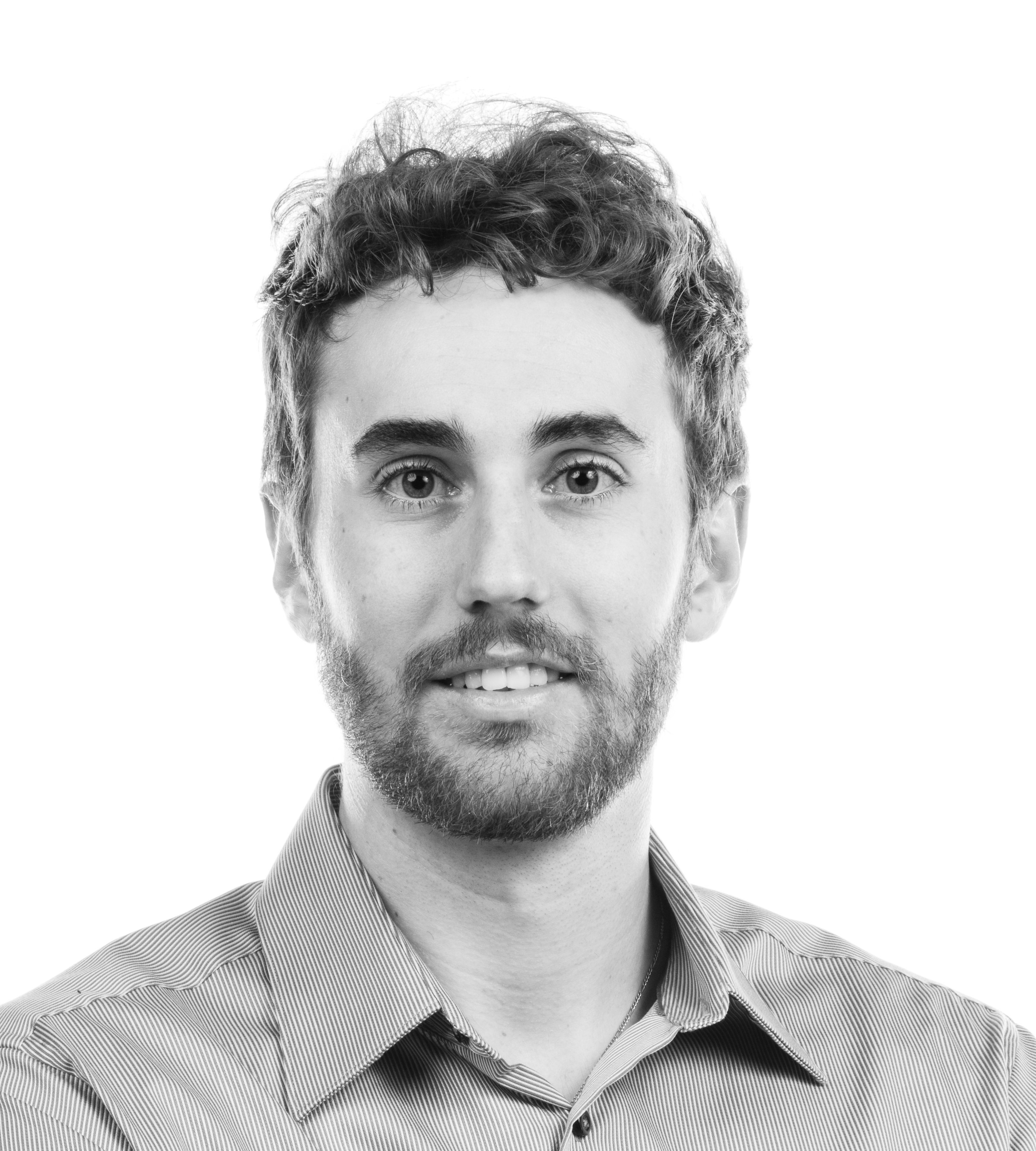}}

\begin{document}
\title{Continuing Isaacson's Legacy: A general metric theory perspective on gravitational memory and the non-linearity of gravity}

\author{ Jann Zosso }

\address{
Center of Gravity, Niels Bohr Institute, Blegdamsvej 17, 2100 Copenhagen, Denmark\\
Albert Einstein Center, Institute for Theoretical Physics, University of Bern, Sidlerstrasse 5, 3012, Bern, Switzerland}

\maketitle\abstracts{
The challenge of defining a physical notion of gravitational waves, together with the associated dynamical degrees of freedom of a gravity theory, is a long-standing problem that famously lead to the discovery the Bondi-Metzner-Sachs (BMS) spacetime symmetry at null infinity and its connection to gravitational memory. Here, we show that the second major contribution to an understanding of waves in gravitation, attributed to the work of Isaacson, equally leads to the inevitable presence of displacement memory, and provides additional understanding of the phenomenon. In particular, the Isaacson viewpoint allows for an efficient method to compute gravitational displacement memory in general metric theories of gravity.}

\section{A Metric Theory Definition of Displacement Memory}

From a modern Wilsonian point of view, any theory of physics at a particular scale is governed by a \textit{principle part} that defines its number and type of dynamical \textit{degrees of freedom} (dofs). In the context of gravity, the principle part corresponds to the classical theory that is required to be well-defined in its entire non-linear form and in particular ought to be fully Ostrogradski stable.~\cite{Zosso:2024xgy} In this language, general relativity (GR), describing two gravitational tensor degrees of freedom, emerges as the natural principle part of a gravity theory with a unique and physical spacetime mertric that couples minimally and universally to matter. A central theory space for any exploration beyond GR is therefore provided by general \textit{metric theories of gravity}, which on top of a universal spacetime metric admit additional dynamical dofs as a characteristic signature of beyond GR effects.~\cite{Zosso:2024xgy}

Due to universal, minimal coupling, any metric theory admits a well-defined geodesic deviation equation that is independent of the equations of motion. In the case of timelike geodesics, it is governed by the six independent components of the electric part of the physical Riemann tensor~\cite{Zosso:2024xgy,maggiore2008gravitational} $\ddot{s}_i=-R_{i 0 j 0} \,s_j$, with $s$ the proper distance between two freely falling test masses. This equation represents the basis of current observations of gravitational radiation. Note that for any finite burst of gravitational radiation from a localized source, there exists the possibility that the proper distance between the test masses in the asymptotic limit does not return to its original value
\begin{equation}\label{eq:defMem}
    \Delta s\neq 0\,.
\end{equation}
Such a finite geodesic deviation that permanently alters the background structure of asymptotic spacetime is known as \textit{gravitational displacement memory}.~\cite{Zeldovich:1974gvh,Christodoulou:1991cr,Blanchet:1992br} 
In fact, gravitational displacement memory is not only a possibility, but is fundamentally present in any emission of gravitational radiation from a localized source.~\cite{Zosso:2024xgy,Christodoulou:1991cr} Following~\cite{Zosso:2024xgy,Heisenberg:2023prj}, this statement can succinctly be understood through the Isaacson approach of defining gravitational waves~\cite{Isaacson_PhysRev.166.1263,Isaacson_PhysRev.166.1272}, as we will now elaborate.

\section{From Isaacson to Memory}

\subsection{The Energy-Momentum of Gravitational Waves and its Back-Reaction}

The definition of gravitational waves in a general spacetime governed by a physical metric $g_{\mu\nu}$ requires a split of the metric 
\begin{equation}
    g_{\mu\nu}=g^L_{\mu\nu}+h^H_{\mu\nu}\,,
\end{equation}
into a background $g^L_{\mu\nu}$ on top of which we whish to describe the propagation of waves $h^H_{\mu\nu}$, together with two key assumptions inherent to the fundamental concept of waves:
\begin{enumerate}[I.)]
    \item Waves are perturbations: $|h^H_{\mu\nu}|\ll|g^L_{\mu\nu}|\sim 1$\,.
    \item There exists a parametric separation in scales of variation between $h^H_{\mu\nu}$ and $g^L_{\mu\nu}$.
\end{enumerate}
The parametric separation in scales of variation that can be imposed in terms of frequency scales~\cite{maggiore2008gravitational} $f_L\ll f_H$, is crucial to ensure the physical uniqueness of the split by providing a natural coarse-graining spacetime average $\langle ...\rangle$~\cite{maggiore2008gravitational,Isaacson_PhysRev.166.1272,Flanagan:2005yc}, such that $\langle g_{\mu\nu}\rangle= g^L_{\mu\nu}$.

Focusing on the vacuum Einstein equations $G_{\mu\nu}=0$ for simplicity, with $G_{\mu\nu}$ the Einstein tensor, it is nowadays a textbook exercise to subsequently arrive at two sets of leading order equations of motion~\cite{Zosso:2024xgy,maggiore2008gravitational,Isaacson_PhysRev.166.1263,Isaacson_PhysRev.166.1272,Flanagan:2005yc,Maccallum:1973gf,misner_gravitation_1973}
\begin{align}
    \phantom{}_{\mys{(1)}}G_{\mu\nu}[g_L,h_H]=0\,,\quad G_{\mu\nu}[g_L]=-\,\big\langle\phantom{}_{\mys{(2)}}G_{\mu\nu}[g_L,h_H]\big\rangle\,.\label{eq:EOMISGR2}
\end{align}
Here, $\phantom{}_{\mys{(i)}}O$ denotes the $i$th order in the perturbative expansion of the operator. The first equation describes a propagation equation for short-scale/high-frequency gravitational waves on top of a generic background $g^L_{\mu\nu}$.
The second equation on the other hand, provides a natural definition for the energy-momentum carried by these gravitational waves
\begin{equation}\label{eq:DefPseudoEMTensorGR}
    t_{\mu\nu}[g_L,h_H]\equiv -\frac{1}{\kappa}\big\langle\phantom{}_{\mys{(2)}}G_{\mu\nu}[g_L,h_H]\big\rangle\,,
\end{equation}
that governs their back-reaction onto the background metric. We defined $\kappa\equiv 8\pi G$, with $G$ the Newton constant. It is important to note, that the spacetime average $\langle ...\rangle$ over the short/fast scales, naturally introduced through the Isaacson approach, is crucial to ensure conservation and gauge invariance of the energy-momentum tensor expression. These statements can be viewed as the major achievement of the work of Isaacson.

\subsection{\label{sec:GWmemFlat}GW Memory in Asymptotically Flat Spacetime}

It took more than half a century to provide a solution to Eqs.~\eqref{eq:EOMISGR2} in the most natural setting to consider gravitational waves: the asymptotically flat spacetime region in the \emph{wave zone} of an isolated source of gravitational waves, described in terms of a source centered Minkowski rest frame \cite{misner_gravitation_1973} $\{t,r,\Omega=(\theta,\phi)\}$. In the appropriate Lorenz gauge, and with the crucial assumption of also considering a perturbation expansion of the background metric $g_{\mu\nu}^L=\eta_{\mu\nu}+h^L_{\mu\nu}$ in terms of distance $r$ from the source~\footnote{More precisely, the expansion is in $\alpha\sim\frac{GM(fGM)^{2/3}}{r}\ll 1$, which in the asymptotic limit is dominated by $r$.}, the Isaacson equations [Eqs.~\eqref{eq:EOMISGR2}] reduce to
\begin{equation}\label{eq:EOMhighfrequM}
    \Box h^H_{\mu\nu}=0\,,\quad  \Box h^L_{\mu\nu}=-2\kappa \, t_{\mu\nu}[\eta,h_H] \,.
\end{equation}
Focusing on the relevant radiative spacial TT components, the general solution of Eqs.~\eqref{eq:EOMhighfrequM} in the asymptotic null limit $r\rightarrow\infty$ at fixed asymptotic retarde time $u\equiv t-r$ reads~\footnote{See also~\cite{PhysRevD.44.R2945,Garfinkle:2022dnm} for a similar computation in alternative frameworks.}~\cite{Heisenberg:2023prj,Heisenberg:2024cjk,Zosso:2024xgy}
\begin{align}\label{eq:DispMemoryGR}
   h_{ij}^{L\text{TT}}(u,r,\Omega)=\,\frac{\kappa}{2\pi r} \int_{-\infty}^udu'\int_{S^2} d^2\Omega'\,\frac{dE_H}{d\Omega'du'}\,\left[\frac{\perp_{ijab}(\Omega)\,\,n'_a n'_b}{1-\,\mathbf{n}'\cdot\mathbf{n}(\Omega)}\right]\,,
\end{align}
where the TT projector can be written as $\perp_{ijab}\,\equiv \,\perp_{ia}\perp_{jb}-\frac{1}{2}\perp_{ij}\perp_{ab}$ with $\perp_{ij}\,\equiv \delta_{ij}-n_in_j$ and $n_i=\partial_i r$. In the wave zone, the GR Isaacson energy flux of high-frequency gravitational waves [Eq.~\eqref{eq:DefPseudoEMTensorGR}] is computed to be
\begin{equation}
    \frac{dE_H}{d\Omega'du'}\equiv r'^2 t_{00}(u',r',\Omega')=\frac{r'^2}{4\kappa}\Big\langle \partial_0 h^H_{\alpha\beta}\partial_0 h_H^{\alpha\beta}\Big\rangle=\frac{r'^2}{2\kappa }\,\Big\langle \dot{h}_{H+}^{2}+\dot{h}_{H\times}^2\Big\rangle\,,
\end{equation}
with $h_+$ and $h_\times$ the well know dynamical dofs of the asymptotic radiation. Again, the physicality of this expression as an energy flux decisively depends on the spacetime averaging $\langle...\rangle$. Indeed, both classically as well as quantum mechanically, the definition of energy carried by wave-like perturbations only makes sense on coarse-grained scales covering at least a few wavelengths.

It can readily be checked, that the solution in Eq.~\eqref{eq:DispMemoryGR} for the leading order background perturbation $h_L$ describes a gravitational displacement memory component as defined in Eq.~\eqref{eq:defMem}. While reaching null infinity as part of the asymptotic radiation, the Isaacson viewpoint implies that this low-frequency component is not part of the emitted gravitational waves, showing that gravitational radiation cannot be reduced to the phenomenon of waves. Moreover, this framework not only provides the final offset of the step-like memory signal governed by the duration of emission $T$ and the scale $f_H$ related to the amount of energy emission, but also gives access to a unique way of defining its smooth time-dependent raise characterized by a frequency cutoff at $f_L\sim 1/T$. Such a natural way of distinguishing between the oscillatory wave- and memory-parts within the radiation is in particular important for future efforts to detect the phenomenon, as typical interferometric observatories are only sensitive to the time-varying component of memory within a given frequency band. Finally, since the energy-momentum carried by emitted gravitational waves will always source such an additional contribution to the background spacetime, any burst of gravitational waves is accompanied by an additional radiative memory signal.~\footnote{Other types of asymptotic energy momentum entering in Eq.~\eqref{eq:EOMISGR2} would also lead to a permanent offset termed linear memory.~\cite{Zosso:2024xgy,Zeldovich:1974gvh,Heisenberg:2024cjk} In this work, we will however focus on the so called non-linear memory as it generally describes the dominant component within compact binary coalescences.~\cite{Christodoulou:1991cr}}
In other words, this mechanism of giving rise to gravitational displacement memory represents a direct measurement of the inevitable non-linearity of gravity.

\section{Gravitational Memory Beyond GR}

The framework to compute and discuss gravitational memory based on Isaacson's basic definitions of gravitational waves not only provides additional insight into gravitational memory within GR, but in particular also allows for a rather straightforward generalization to metric theories beyond GR. Indeed, based on Isaacson's work, it was possible to develop a novel method for computing gravitational displacement memory in metric theories of gravity, giving access to entirely new avenues to test the foundations of metric theories and the characteristic non-linearity of gravitation.~\cite{Heisenberg:2023prj,Zosso:2024xgy,Heisenberg:2024cjk,Heisenberg:2025tfh} 

As an example, the first main result of this program was a computation of the complete memory formula in massless Horndeski theory~\cite{Horndeski:1974wa}, representing the most general covariant metric theory with additional scalar field dof $\varphi$ up to two powers of derivatives per fields.
In the notation of~\cite{Zosso:2024xgy}, the corresponding tensor displacement memory formula is given by~\cite{Zosso:2024xgy,Heisenberg:2023prj}
\begin{align}\label{eq:DispMemoryHorndeski}
   h_{ij}^{L\text{TT}}(u,r,\Omega)=\,\frac{1}{4\pi r} \int_{-\infty}^udu'\int_{S^2} d^2\Omega'\,r'^2\Big\langle \dot{h}_{H+}^2+\dot{h}_{H\times}^2+\rho^2\,\dot{\varphi}_H^2\Big\rangle\,\left[\frac{\perp_{ijab}(\Omega)\,\,n'_a n'_b}{1-\,\mathbf{n}'\cdot\mathbf{n}(\Omega)}\right]\,,
\end{align}
where the coefficient $\rho$ in terms of general functionals $G_i$ that define Horndeski theory reads
\begin{equation}
    \rho=\sqrt{3\,\sigma^2+\frac{(\bar G_{2,X}-2\,\bar G_{3,\Phi})}{\bar G_4}}\,,\quad  \sigma= \frac{\bar G_{4\Phi}}{\bar G_{4}}\,.
\end{equation}
This formula represents the first step towards an exploration of the non-linear strong field dynamics around the merger of a black hole binary with the additional component of the memory. Moreover, the fact that the tensor memory is sensitive to all three propagating dofs of the theory might promise a future application of memory as a reliable verification channel for one of the cleanest smoking gun signals beyond GR.

\section*{Acknowledgments}

JZ is supported by funding from the Swiss National Science Foundation
grant 222346. The Center of Gravity is a Center of Excellence funded by the Danish National Research Foundation under grant No. 184.

\section*{References}
\bibliography{moriond}

\begin{thebibliography}{10}

\bibitem{Zosso:2024xgy}
Jann Zosso.
\newblock {\em {Probing Gravity - Fundamental Aspects of Metric Theories and their Implications for Tests of General Relativity}}.
\newblock PhD thesis, Zurich, ETH, 2024.

\bibitem{maggiore2008gravitational}
Michele Maggiore.
\newblock {\em Gravitational Waves: Volume 1: Theory and Experiments}.
\newblock Oxford University Press, 10 2007.

\bibitem{Zeldovich:1974gvh}
Y.~B. Zel'dovich and A.~G. Polnarev.
\newblock Radiation of gravitational waves by a cluster of superdense stars.
\newblock {\em Sov. Astron.}, 18:17, 1974.

\bibitem{Christodoulou:1991cr}
D.~Christodoulou.
\newblock Nonlinear nature of gravitation and gravitational wave experiments.
\newblock {\em Phys. Rev. Lett.}, 67:1486--1489, 1991.

\bibitem{Blanchet:1992br}
Luc Blanchet and Thibault Damour.
\newblock Hereditary effects in gravitational radiation.
\newblock {\em Phys. Rev. D}, 46:4304--4319, 1992.

\bibitem{Heisenberg:2023prj}
Lavinia Heisenberg, Nicol\'as Yunes, and Jann Zosso.
\newblock Gravitational wave memory beyond general relativity.
\newblock {\em Phys. Rev. D}, 108(2):024010, 2023.

\bibitem{Isaacson_PhysRev.166.1263}
Richard~A. Isaacson.
\newblock Gravitational radiation in the limit of high frequency. i. the linear approximation and geometrical optics.
\newblock {\em Phys. Rev.}, 166:1263--1271, Feb 1968.

\bibitem{Isaacson_PhysRev.166.1272}
Richard~A. Isaacson.
\newblock Gravitational radiation in the limit of high frequency. ii. nonlinear terms and the effective stress tensor.
\newblock {\em Phys. Rev.}, 166:1272--1280, Feb 1968.

\bibitem{Flanagan:2005yc}
Eanna~E. Flanagan and Scott~A. Hughes.
\newblock The basics of gravitational wave theory.
\newblock {\em New J. Phys.}, 7:204, 2005.

\bibitem{Maccallum:1973gf}
Malcolm A.~H. Maccallum and A.~H. Taub.
\newblock The averaged lagrangian and high-frequency gravitational waves.
\newblock {\em Commun. Math. Phys.}, 30:153--169, 1973.

\bibitem{misner_gravitation_1973}
Charles~W. Misner, K.~S. Thorne, and J.~A. Wheeler.
\newblock {\em Gravitation}.
\newblock W. H. Freeman, San Francisco, 1973.

\bibitem{PhysRevD.44.R2945}
Alan~G. Wiseman and Clifford~M. Will.
\newblock Christodoulou's nonlinear gravitational-wave memory: Evaluation in the quadrupole approximation.
\newblock {\em Phys. Rev. D}, 44:R2945--R2949, Nov 1991.

\bibitem{Garfinkle:2022dnm}
David Garfinkle.
\newblock Gravitational wave memory and the wave equation.
\newblock {\em Class. Quant. Grav.}, 39(13):135010, 2022.

\bibitem{Heisenberg:2024cjk}
Lavinia Heisenberg, Guangzi Xu, and Jann Zosso.
\newblock {Unifying ordinary and null memory}.
\newblock {\em JCAP}, 05:119, 2024.

\bibitem{Heisenberg:2025tfh}
Lavinia Heisenberg, Benedetta Rosatello, Guangzi Xu, and Jann Zosso.
\newblock {Constraining Superluminal Einstein-\AE{}ther Gravity through Gravitational Memory}.
\newblock 5 2025.

\bibitem{Horndeski:1974wa}
Gregory~Walter Horndeski.
\newblock Second-order scalar-tensor field equations in a four-dimensional space.
\newblock {\em Int. J. Theor. Phys.}, 10:363--384, 1974.

\end{thebibliography}


\end{document}